

\input phyzzx.tex
\overfullrule0pt

\def\ie{{\it i.e.}}
\def\etal{{\it et al.}}
\def\half{{\textstyle{1 \over 2}}}

\def\bold#1{\setbox0=\hbox{$#1$}%
     \kern-.025em\copy0\kern-\wd0
     \kern.05em\copy0\kern-\wd0
     \kern-.025em\raise.0433em\box0 }
\Pubnum={VAND-TH-94-19}
\date={August 1994}
\pubtype{}
\titlepage


\vskip1cm
\title{\bf Renormalizing Coupled Scalars with a Momentum Dependent
Mixing Angle in the MSSM ${}^{*}$}
\vskip .1in
\author{Marco Aurelio D\'\i az }
\vskip .1in
\centerline{Department of Physics and Astronomy}
\centerline{Vanderbilt University, Nashville, TN 37235}
\vskip .2in

\centerline{\bf Abstract}
\vskip .1in

The renormalization of a system of coupled scalars fields is analyzed.
By introducing a momentum dependent mixing
angle we diagonalize the inverse propagator matrix at any momentum
$p^2$. The zeros of the inverse propagator matrix, \ie, the physical masses,
are then calculated
keeping the full momentum dependence of the self energies.
The relation between this method and others previously published
is studied. This idea is applied to the one-loop renormalization of
the CP-even neutral Higgs sector of the Minimal Supersymmetric Model,
considering top and bottom quarks and squarks in the loops.

\vskip 2cm
\noindent * Presented in the Eighth Meeting of the Division of
Particles and Fields of the American Physical Society ``DPF'94'',
The University of New Mexico,
Albuquerque NM, August 2-6, 1994.

\vfill

\endpage

\voffset=-0.2cm

\REF\michael{M. Capdequi-Peyran\`ere and M. Talon, {\sl Il Nuovo
   Cimento} {\bf 75}, 205 (1983).}

Several years ago, Capdequi-Peyran\`ere and Talon\refmark\michael
studied the wave function renormalization of coupled systems of
scalars, fermions, and vectors. Their approach includes conventional
mass counterterms and wave function renormalization plus a
non-conventional field ``rotation'' that allow them to
impose no mixing between the states at different scales, those scales
being the masses of the different states. Here we
generalize this idea and, at the same time, explain the nature of this
field ``rotation''.

Similarly to ref.~[\michael], consider the bare lagrangian corresponding
to a system of two scalars:
$${\cal L}_b=\half\chi_{1b}(p^2-m_{1b}^2)\chi_{1b}+\half\chi_{2b}(p^2-
m_{2b}^2)\chi_{2b}-\chi_{1b}m_{12b}^2\chi_{2b}\eqn\Lbare$$
If we denote by $-iA_{ij}^{\chi}(p^2)$, $i,j=1,2$, the sum of the
one-loop Feynman graphs contributing to the two point
functions and, after shifting the bare masses by $m_{ib}^2\rightarrow
m_i^2-\delta m_i^2$, $i=1,2,12$, and the fields by $\chi_{ib}\rightarrow
(1+\half\delta Z_i)\chi_i$, the effective lagrangian is
$${\cal L}_{eff}=\half(\chi_1,\chi_2){\bold\Sigma^{\bold\chi}}
{\chi_1\choose\chi_2}\eqn\Leffect$$
with
$$\eqalign{\Sigma_{11}^{\chi}(p^2)=&p^2-m_1^2+(p^2-m_1^2)
\delta Z_1+\delta m_1^2-A_{11}^{\chi}(p^2)\cr
\Sigma_{22}^{\chi}(p^2)=&p^2-m_2^2+(p^2-m_2^2)
\delta Z_2+\delta m_2^2-A_{22}^{\chi}(p^2)\cr
\Sigma_{12}^{\chi}(p^2)=&-m_{12}^2-\half m_{12}^2(\delta Z_1+\delta Z_2)
+\delta m_{12}^2-A_{12}^{\chi}(p^2)\cr}\eqn\sigmachielem$$
Although it is not a necessary assumption, in order to compare more
easily with ref.~[\michael], we assume for the moment that the two
scalars are diagonal at tree level, \ie, $m_{12}^2=0$. In this case,
if we want the pole of the propagators to be the physical masses
with a residue equal to unity, the two mass counterterms and the two
wave function renormalization are fixed through the relations
$$\eqalign{&\delta m_1^2=A_{11}^{\chi}(m_1^2),\qquad\delta m_2^2=A_{22}
^{\chi}(m_2^2)\cr
&\delta Z_1={A'}_{11}^{\chi}(m_1^2),\qquad\delta Z_2={A'}_{22}^{\chi}
(m_2^2)\cr}\eqn\counterterms$$
where the prime denote the derivative with respect to the argument.
We may want to fix the $\delta m_{12}^2$ counterterm by imposing
no mixing between $\chi_1$ and $\chi_2$ at a given scale, for example
at $p^2=m_1^2$. In this case, the off diagonal element of the inverse
propagator matrix is
$$\Sigma_{12}^{\chi}(p^2)=\delta m_{12}^2-A_{12}^{\chi}(p^2)=
A_{12}^{\chi}(m_1^2)-A_{12}^{\chi}(p^2)\equiv-\tilde A_{12}^{\chi}
(p^2)\eqn\sigmaoffdiag$$
At this point all the counterterms are fixed, and since
$\tilde A_{12}^{\chi}(p^2)$ is zero only at $p^2=m_1^2$,
the two fields are not decoupled at a different scale.
This is the motivation for the authors in ref.~[\michael] to define
the unconventional wave function renormalization $\chi_{1b}\rightarrow
(1-\alpha_1)\chi_1-\beta_1\chi_2$ and $\chi_{2b}\rightarrow(1-\alpha_2)
\chi_2-\beta_2\chi_1$ instead of $\chi_{1b}\rightarrow(1+\half\delta
Z_i)\chi_i$ we use here. Setting to one the residue of the pole of
each propagator they find
$$\alpha_i=-\half\delta Z_i=-\half{A'}_{ii}^{\chi}(m_1^2),\qquad i=1,2
\eqn\alfas$$
and imposing no mixing between the two fields at $p^2=m_1^2$ and {\it
also} at $p^2=m_2^2$ they get
$$\beta_1={{A_{12}^{\chi}(m_2^2)}\over{m_1^2-m_2^2}},\qquad\beta_2=
{{A_{12}^{\chi}(m_1^2)}\over{m_2^2-m_1^2}}\eqn\betas$$
In the language we are using here, the later is equivalent
to perform a ``rotation'' (we already know that it is not a field rotation,
it is just a wave function renormalization that mixes the two fields)
to the inverse propagator matrix in the following way
$${\bold{\Sigma^{\chi}}}\longrightarrow\left[\matrix{1&\beta_2\cr
\beta_1&1\cr}\right]{\bold{\Sigma^{\chi}}}\left[\matrix{1&\beta_1\cr
\beta_2&1\cr}\right]\eqn\betarotation$$

\REF\thesis{M.A. D\'\i az, Ph.D. thesis, preprint SCIPP-92/13
   (1992).}

Here we propose a modification of this procedure. If we define
a momentum dependent mixing angle we will be able to diagonalize the
inverse propagator matrix at {\it any} momentum\refmark\thesis. Considering
the already finite inverse propagator matrix elements in
eq.~\sigmachielem,
we define a momentum dependent mixing angle $\alpha(p^2)$ by
$$\tan[2\alpha(p^2)]={{2\Sigma_{12}^{\chi}(p^2)}\over
{\Sigma_{11}^{\chi}(p^2)-\Sigma_{22}^{\chi}(p^2)}}
\eqn\defalpha$$
The matrix ${\bold\Sigma^{\bold\chi}}(p^2)$ is diagonalized at
any momentum $p^2$ by a rotation defined by the angle
$\alpha(p^2)$
$${\bold\Sigma^{\chi}}\longrightarrow\left[\matrix{
c_{\alpha}&s_{\alpha}\cr-s_{\alpha}&c_{\alpha}\cr}\right]{\bold
\Sigma^{\chi}}\left[\matrix{c_{\alpha}&-s_{\alpha}\cr s_{\alpha}&
c_{\alpha}\cr}\right]\eqn\rotsigma$$
where $s_{\alpha}$ and $c_{\alpha}$ are sine and cosine of the momentum
dependent mixing angle $\alpha(p^2)$. Considering eqs.~\sigmachielem,
\sigmaoffdiag, and~\defalpha\ we find in first approximation
$$s_{\alpha}(p^2)\approx{{\tilde A_{12}^{\chi}(p^2)}\over{m_2^2-m_1^2}},
\qquad c_{\alpha}\approx1\eqn\salphaappro$$
making evident the relation with the previous method.

There is a fundamental difference between our approach and the one in
ref.~[\michael]: we are rotating an already finite inverse propagator
matrix with a rotation matrix defined by a finite momentum dependent
mixing angle, on the contrary, in ref.~[\michael] the ``rotation'' is
in fact a wave function renormalization that mixes the two fields
and the ``rotation'' matrix elements $\beta_1$ and $\beta_2$ are
infinite. A momentum dependent field rotation given by
eqs.~\defalpha\ and~\rotsigma\ is the only way to diagonalize
the inverse propagator matrix at any scale. On the other hand, the
only momentum independent way to diagonalize this matrix at two
scales ($p^2=m_1^2$ and $p^2=m_2^2$) is with the field ``rotation''
in eqs.~\betas\ and~\betarotation. By contrast, the conventional
renormalization of this matrix will allow us to diagonalize it at only
one scale. According to our example in eq.~\sigmaoffdiag, that scale is
$p^2=m_1^2$, and any further momentum independent rotation of the fields
will diagonalize the inverse propagator matrix at a different scale, for
example at $p^2=m_2^2$, by using $\alpha(p^2=m_2^2)$ in
eq.~\salphaappro,
but spoiling the previous diagonalization (at $p^2=m_1^2$).

Using a momentum dependent mixing angle in this way is an alternative
to define a counterterm for this angle. In fact, the renormalization
procedure is carried out in the unrotated basis and no mixing angle is
defined at that level. Similarly, instead of renormalizing couplings
of the rotated fields to other particles, we renormalize couplings
of the unrotated fields to those particles and after that we rotate
by an angle $\alpha(p^2)$, where $p^2$ is the typical scale of the process,
for example, $p^2=m_i^2$ if the rotated field $\chi_i$ is on-shell.
Usually, working out the radiative corrections in the unrotated basis
implies one extra advantage, and that is the simplicity of the
Feynman rules. In the following we will illustrate these ideas by
renormalizing the CP-even neutral Higgs masses of the Minimal
Supersymmetric Model (MSSM).

\REF\ChargedH{J.F. Gunion and A. Turski, {\sl Phys. Rev. D} {\bf 39},
   2701 (1989); {\bf 40}, 2333 (1989); A. Brignole, J. Ellis,
   G. Ridolfi, and F. Zwirner,
   {\sl Phys. Lett. B} {\bf 271}, 123 (1991); M. Drees
   and M.M. Nojiri, {\sl Phys. Rev. D} {\bf 45}, 2482 (1992);
   A. Brignole, {\sl Phys. Lett. B} {\bf 277}, 313 (1992);
   P.H. Chankowski, S. Pokorski, and J. Rosiek,
   {\sl Phys. Lett. B} {\bf 274}, 191 (1992);
   M.A. D\'\i az and H.E. Haber, {\sl Phys. Rev. D}
   {\bf 45}, 4246 (1992).}
\REF\neutral{M.S. Berger, {\it Phys. Rev. D} {\bf 41}, 225 (1990);
   H.E. Haber and R. Hempfling,
   \sl Phys. Rev. Lett. {\bf 66}, \rm 1815 (1991);
   Y. Okada, M. Yamaguchi and T. Yanagida,
   \sl Prog. Theor. Phys. {\bf 85}, \rm 1 (1991);
   J. Ellis, G. Ridolfi and F. Zwirner, \sl Phys. Lett.
   {\bf B257}, \rm 83 (1991);
   R. Barbieri, M. Frigeni, F. Caravaglios, \sl Phys. Lett. {\bf B258},
   \rm 167 (1991);
   Y. Okada, M. Yamaguchi and T. Yanagida,
   \sl Phys. Lett. {\bf B262}, \rm 54 (1991);
   J.R. Espinosa and M. Quir\' os, {\it Phys. Lett. B} {\bf 266},
   389 (1991);
   A. Yamada, \sl Phys. Lett. {\bf B263}, \rm 233 (1991);
   J. Ellis, G. Ridolfi and F. Zwirner, {\sl Phys.
   Lett.} {\bf B262}, {\rm 477 (1991)};
   M. Drees and M.M. Nojiri, {\it Phys. Rev. D} {\bf 45}, 2482 (1992);
   R. Barbieri and M. Frigeni, {\sl Phys. Lett.} {\bf
   B258}, 395 (1991);
   A. Brignole, {\it Phys. Lett. B} {\bf 281}, 284 (1992);
   M.A. D\'\i az and H.E. Haber, {\it Phys. Rev. D}
   {\bf 46}, 3086 (1992).}

The radiative corrections to the Higgs masses in the MSSM have
been studied by many authors in the
last three years, using different techniques and focusing
in different particles and processes. It was established the
theoretical convenience of parametrizing the Higgs sector
through the CP-odd Higgs mass ($m_A$) and the ratio of the
vacuum expectation values of the two Higgs doublets ($\tan\beta
=v_2/v_1$). The radiative corrections to the charged Higgs mass
were found to be small\refmark{\ChargedH},
growing as $m_t^2$, unless there is
an appreciable mixing in the squark mass matrix: in that case
a term proportional to $m_t^4$ is non-negligible.
The corrections to the CP-even Higgs masses are large and
grow as $m_t^4$, and have profound consequences in the phenomenology
of the Higgs sector\refmark{\neutral}.

\REF\diazalfa{M.A. D\'\i az, Report Number VAND-TH-94-16, in progress.}
\REF\diazHdecay{M.A. D\'\i az, {\sl Phys. Rev. D} {\bf 48}, 2152
   (1993).}

We renormalize the CP-even Higgs sector of the MSSM working in an
on-shell type of scheme\refmark\diazalfa, where the physical masses
of the gauge bosons $m_Z$ and $m_W$, and of the CP-odd Higgs $m_A$
correspond to the pole of the propagators. The parameter $\tan\beta$
is defined through the renormalization of the CP-odd Higgs vertex
to a pair of charged leptons\refmark\diazHdecay. The electric charge
is defined through the photon coupling to a positron-electron pair.
We also set the residue
of the photon and CP-odd Higgs equal to unity, and impose no mixing
between the photon and the $Z$ boson at zero momentum.

\REF\CDF{F. Abe \etal, CDF Collaboration, Fermilab Pub--94/097--E
  	(1994).}

\FIG\figi{Momentum dependent mixing angle $\alpha$ (in degrees)
as a function of $\sqrt{p^2}$, for the five different values of the
CP-odd Higgs mass $m_A=500, 130, 115, 80, 50$ GeV. We consider the
CDF preferred value for the top quark mass and no squark mixing.
All the other soft supersymmetry breaking mass terms are equal to
1 TeV. The two crosses over each curve correspond to the masses
of the two CP-even Higgs bosons.}

In Fig. \figi\ we plot the mixing angle $\alpha(p^2)$ as a function of the
momentum $p^2$ for different values of the CP-odd Higgs mass $m_A$.
For the top quark mass we take the CDF preferred value $m_t=174$
GeV\refmark\CDF. We take $\tan\beta=25$ and all the squark soft
supersymmetry breaking mass terms equal to 1 TeV. We consider no-mixing
in the squark mass matrices ($\mu=A_U=A_D=0$). The two crosses in each
curve represent the masses of the two CP-even Higgs bosons. We see that
the angle $\alpha$ at the scale $m_h$ is very close to the one at the scale
$m_H$, even for large splitting between these two masses, like in the case
$m_A=500 GeV$. In this later situation, the angle $\alpha\rightarrow0$,
consistent with the decoupling of the heavy Higgs (in general
$\alpha\rightarrow\beta-\pi/2$, and in this case, $\beta\approx\pi/2$).

\FIG\figii{Coefficient $\sin(\beta-\alpha)$, the MSSM/SM ratio of
the $ZZh$ vertex, as a function of $\tan\beta$. We show the tree level
value (dots), the improved value with the leading $m_t^4$ term
(dotdash), and the value calculated with $\alpha(p^2)$ for two different
choices of the squark mixing parameters and two different scales:
$p^2=m_H^2$ (dashes) and $p^2=m_h^2$ (solid).}

An important mechanism for the production of the neutral Higgs bosons
in $e^+e^-$ colliders is the brehmsstrahlung of a Higgs by a $Z$ gauge boson.
Relative to the coupling of the SM higgs to two $Z$ bosons,
the $ZZh$ coupling is $\sin(\beta-\alpha)$. We plot this parameter in
Fig. \figii\ as a function of $\tan\beta$. We contrast the tree level answer
(dotted line) and the improved version (dashed line) defined by
$$\tan 2\alpha={{(m_A^2+m_Z^2)s_{2\beta}}\over{
(m_A^2-m_Z^2)c_{2\beta}+\Delta_t}}, \quad {\rm with}\qquad
\Delta_t={{3g^2m_t^4}\over{16\pi^2m_W^2s_{\beta}^2}}\,\ln\,
{{m_{\tilde t_1}^2m_{\tilde t_2}^2}\over{m_t^4}},\eqn\alfadeltat$$
with the parameter calculated with the momentum dependent mixing angle
$\alpha(p^2)$. In this later case, we plot the result using $\alpha(p^2)$
evaluated at the two well motivated scales given by the masses of the
two CP-even Higgs, and we find small differences between these two
scales. However, important differences are found with the tree level
and the improved cases, indicating that this effect may be important
in the search of the Higgs boson at LEP.

\vskip .5cm
\centerline{\bf ACKNOWLEDGMENTS}
\vskip .5cm

It is greatly appreciated the help of Howard E. Haber in the early
stages of this work.
Useful comments by Tonnis ter Veldhuis,
and Thomas J. Weiler are acknowledged.
This work was supported by the U.S. Department of Energy, grant No.
DE-FG05-85ER-40226, and by the Texas National Research Laboratory
(SSC) Commission, award No. RGFY93-303.

\refout
\figout
\end